\documentclass[conference]{IEEEtran}
\IEEEoverridecommandlockouts
\usepackage{cite}
\usepackage{amsmath,amssymb,amsfonts}
\usepackage{algorithmic}
\usepackage{graphicx}
\usepackage{textcomp}
\usepackage{xcolor}
\usepackage{subfigure}
\usepackage{array}
\usepackage{booktabs}
\usepackage{multirow}
\usepackage{multicol}
\usepackage{threeparttable}
\def\BibTeX{{\rm B\kern-.05em{\sc i\kern-.025em b}\kern-.08em
    T\kern-.1667em\lower.7ex\hbox{E}\kern-.125emX}}
\begin{document}

\title{PREAMBLE and IMRECEIVING for Improved Large Message Handling in libp2p GossipSub\\
}

\author{\IEEEauthorblockN{1\textsuperscript{st} Muhammad Umar Farooq}
\IEEEauthorblockA{\textit{Vac Research} \\
\textit{Institute of Free Technology (IFT),}\\
Singapore \\
farooq@status.im}
\and
\IEEEauthorblockN{2\textsuperscript{nd} Daniel Kaiser}
\IEEEauthorblockA{\textit{Vac Research} \\
\textit{Institute of Free Technology (IFT),}\\
Singapore \\
danielkaiser@status.im}
}

\maketitle

\begin{abstract}
Large message transmissions in libp2p GossipSub lead to longer than expected network-wide message dissemination times and very high bandwidth utilization. This article identifies key issues responsible for this behavior and proposes modifications to the protocol for transmitting large messages. These modifications preserve the GossipSub resilience and fit well into the current algorithm. The proposed changes are rigorously evaluated for performance using the shadow simulator. Results reveal that the suggested changes reduce bandwidth utilization by up to 61\% and message dissemination time by up to 35\% under different traffic conditions.      
\end{abstract}


\section{Introduction}
Message dissemination mechanisms in unstructured peer-to-peer (P2P) networks elucidate the network's performance, scalability, and robustness. Flooding \cite{lin1999gossip, ai2006efficient, margariti2013study}, for instance, is the most straightforward and robust method that can achieve minimum message dissemination latency and increased resilience against failures as the message propagates through all (including the optimal) paths. But, very high bandwidth utilization compromises scalability in flooding. On the other hand, probabilistic message forwarding \cite{escalante2005rng, tumas2022probabilistic} may result in lower bandwidth utilization and better scalability by putting total coverage and network resilience at risk. At the same time, tree-based/hierarchical approaches \cite{srivatsa2004scaling, duan2017two, qiu2022geography} may maintain moderate bandwidth and message dissemination latency. However, they involve a higher overlay management complexity attributed to the additional coordination and control needed for perpetual backbone maintenance. Incorporating the publish-subscribe (PubSub) \cite{eugster2003many, ipfsblog2017pubsub, git2020pubsub} paradigm into message dissemination mechanisms helps achieve scalability using content-centered message dissemination. In PubSub, peers subscribe to their topics of interest and receive messages that strictly belong to the subscribed topics without needing any direct connection between the publisher and the subscriber. This approach limits the scope of message dissemination and enhances scalability. A large number of PubSub variants are available. For instance, in FloodSub \cite{benet2016floodsub}, nodes send a message to all known peers interested in a specific topic. On the other hand, EpiSub \cite{git2018episub} segregates subscribed peers into active and passive groups, and the message is only propagated through active peers to maintain relatively lower network-wide bandwidth utilization.   

GossipSub \cite{vyzovitis2020gossipsub} is one of the most widely adopted PubSub protocols. It strives to offer a FloodSub-like message dissemination latency and network resilience, while maintaining moderate bandwidth utilization. 
Hosts in GossipSub maintain a full-message mesh (comprising $D$ peers, randomly selected from the pool of known peers subscribed to a specific topic) for limited flooding and a metadata-only mesh (employing peers that are not part of the full-message mesh) for gossip emission\footnote{This article uses the terms 'full-message mesh' and 'mesh' interchangeably. Similarly, 'metadata-only mesh' and 'gossip mesh' are interchangeable.}. 
On receiving a message, a GossipSub peer forwards it to mesh members (excluding the ones that delivered this message) and announces the message availability to the gossip mesh during heartbeat intervals. The gossip mainly includes IHAVE announcements carrying IDs of seen messages. 
IHAVE announcements provide additional resilience against non-conforming peers and in the presence of network partitions. On receiving an IHAVE announcement about an unseen message ID (msgID), the receiver sends an IWANT request to the sender. 
In GossipSub v1.0  \cite{vyzo2020gossipsubv1_0}, replying to IWANT requests is mandatory, and there is no limit on the number of unique IWANT requests a peer can make, which can potentially overwhelm many nodes. 
Security hardening extensions in GossipSub v1.1 \cite{vyzo2021gossipsubv1_1} use peer scoring and budget to cater to these problems. Under this arrangement, peers maintain a local score for each known host. This score rises for events like early message delivery, IWANT response, etc., and lowers for events like message validation failure, budget violation, etc. Peer score helps maintain a better mesh as we can replace low-scoring peers. 
Replying to IWANT requests is optional in GossipSub v1.1, which encourages any peer to make multiple IWANT requests for the same message. 
It may adversely impact applications that require timely dissemination of large messages, ranging from several kilobytes to multiple megabytes. This article considers messages exceeding 50 KB as large messages. Much larger sizes are common in practice; for example, Ethereum supports data blobs with a theoretical maximum block size of 32 MB \cite{fulldas2024}.
It is important to note that GossipSub does not consider message size in forwarding (and IWANT message) decisions. This oversight leads to unexpected spikes in message dissemination time and bandwidth utilization.
IDONTWANT messages are introduced in GossipSub v1.2 \cite{nashat2023gossipsubv1_2} to curtail redundant transmissions of large messages. On receiving a message bigger than the specified threshold, peers announce successful reception by sending IDONTWANT messages to the remaining mesh members, who, in response, refrain from resending the same message. This mechanism helps minimize duplicates to some extent. However, additional efforts are needed to eliminate duplicates.

This article identifies key factors that contribute to high bandwidth utilization and prolonged dissemination times for large messages in GossipSub. We propose simple yet effective solutions to mitigate these issues. It is important to note that the proposed changes take effect only when the message size exceeds a specified threshold. The primary contributions of this article are as follows:

\begin{enumerate}
    \item We show that peers are unaware of ongoing message receptions during prolonged transfer times for large messages, which leads to an increase in duplicate messages. We introduce a new message called PREAMBLE to facilitate quicker identification of incoming messages.

    \item We introduce a new message called IMRECEIVING. IMRECEIVING message notifies mesh members about ongoing message reception, asking them to refrain from resending the same message. This prompt notification significantly reduces the number of duplicates and decreases message dissemination time for large messages. 

    \item We use PREAMBLE to identify ongoing message receptions and leverage this information to reduce the number of IWANT requests for large messages. This modification helps reduce message dissemination latency by lowering the workload for initial message senders.
    
    \item We investigate the tradeoffs and benefits of reducing message transmissions to only a subset of mesh members while ensuring resilience by sending immediate IHAVE announcements to the remaining mesh members. 
\end{enumerate} 

We demonstrate that the proposed improvements yield noticeable performance gains that become increasingly significant as the size or the number of messages increases.
The rest of the article is organized as follows: Section \ref{ProblemDescription}  presents the problem description and reviews recent research efforts addressing these challenges. Section \ref{Proposed} discusses the proposed modifications to the GossipSub protocol. Section \ref{Results} provides performance evaluations, and section \ref{Conclusion} concludes the article.

\section{Problem Description and Background} \label{ProblemDescription}
This section explains the main challenges associated with disseminating large messages in GossipSub and summarizes recent research efforts aimed at addressing these challenges. For comprehensive insights into recent advancements in scalability and large message handling in unstructured P2P networks, readers are encouraged to refer to our recent work \cite{farooq2025staggering}.

Considering same link latency and data rate, a message dissemination to full-message mesh concludes in $\tau_D \approx (D \times \tau_{tx}) + \tau_p$ time, where $\tau_{tx} = \frac{S}{R}$, with $S$, $R$, and $\tau_p$ being the message size, data rate, and link latency, respectively. This implies that a tenfold increase in message size results in an eightyfold rise in $D \times \tau_{tx}$ for a mesh with $D = 8$. This enormous rise in message transmission times is not considered in the message forwarding decisions in GossipSub, which results in various performance issues: 

\begin{enumerate}
\item Sending a message to N peers involves approximately $\lceil \log_D(N) \rceil$ rounds, each lasting around $\tau_D$ time. During each round, there are roughly $(D-1)^{X-1} \times D$ transmissions, where $X$ represents the round number. Publisher transmitting to a higher number of peers (floodpublish) can theoretically reduce the network-wide message dissemination time by increasing the transmissions in each round to $(D-1)^{X-1} \times (F+D)$, where $F$ represents the additional number of peers included in floodpublish. This arrangement works fine for relatively small (or moderate) message sizes. However, a smaller congestion window ($C_{wnd}$) at less frequently used links (cold connections) may noticeably increase message dissemination time to such peers. As a result, performance improvement strategies such as floodpublish, IWANT replies, and quicker mesh readjustments may not prove effective when forwarding large messages.

\item $\tau_D$ accumulates at each hop, significantly increasing the time required for network-wide dissemination of messages. This implies that an increase in $\tau_D$ also substantially raises the associated store-and-forward delay. We can estimate network-wide dissemination time as $\tau_{N} \approx \tau_D \times h$, where h represents the number of hops in the longest path. Considering cumulative transmit delay as $\delta_{tx} \approx \frac{D \times S}{R} \times h$, we get $\tau_N \approx \delta_{tx} + (\tau_p \times h)$. Our previous work \cite{farooq2025staggering} demonstrates that partitioning a message in $n$ fragments reduces cumulative transmit delay to $\delta_{tx} \approx \frac{D \times S}{R} \times \frac{2h-1}{n}$, which effectively minimizes store-and-forward delay. The same work also uses message staggering to concentrate the sender's bandwidth for one transmission at a time. These sequential transmissions reduce the time required for a transmission round to $\tau_1$. The number of peers receiving the message during each round increases by $2^X\ \forall\ X \in \{0, D-1\}$ and by $\sum_{k=X-D}^{X-1} \lambda_k\ \forall\ X \geq D$, where $X$ represents message transmission round, and $\lambda_k$ represents the number of peers that received the message in round $k$. 

\item A longer $\tau_D$ increases the likelihood that heartbeat interval(s) will elapse before the message is transmitted. IHAVE announcements from early message receivers may fetch many IWANT requests, thereby increasing the workload on peers situated along the optimal forwarding paths. Notably, many of these IWANT requests may be duplicates or originate from peers already receiving the same message from their mesh members.

\item An increase in $\tau_D$ also raises the likelihood of simultaneous redundant transmissions to the same peer from different senders. Minimizing duplicates by reducing $D$ compromises the network resilience and may also increase network-wide dissemination time. 
IDONTWANT messages \cite{nashat2023gossipsubv1_2} help eliminate duplicates without compromising resilience to achieve reasonable bandwidth improvements. However, studies reveal that further investigations are needed to achieve higher performance improvements from IDONTWANT messages \cite{Yiannis2024num_dup}. 
In \cite{csaba2025pppt}, authors use a hybrid push-pull mechanism to reduce duplicates and induce additional delay when relaying a message to mesh members, thereby improving the effectiveness of IDONTWANT messages. 
Similarly, the works in \cite{pop2024pr617, pop2024pr653} adopt a pull-based operation to eliminate duplicates. However, these methods trade off latency to minimize bandwidth utilization.

\item A busy peer with a sizeable outgoing message load will enqueue (or simultaneously transfer) new messages. This approach prioritizes already-scheduled messages, introducing a significant initial delay to the locally (or newly) published messages. Lack of adaptiveness and standardization in outgoing message prioritization are key factors that can lead to noticeable inconsistency in message dissemination time at each hop, even in similar network conditions. The use of QUIC stream prioritization \cite{iyengar2021rfc} or similar approaches can enable quicker dissemination of high-impact messages, such as IDONTWANTs and newly published messages.

\end{enumerate}

We already explored the use of message staggering and fragmentation with IDONTWANT messages in \cite{farooq2025staggering} to minimize the store-and-forward delay inherent to large message transmissions in GossipSub. This article focuses on reducing duplicates, which also lowers the overall message dissemination time by minimizing the workload of peers.

\begin{figure*}[htb]
\centerline{\includegraphics[width=\linewidth]{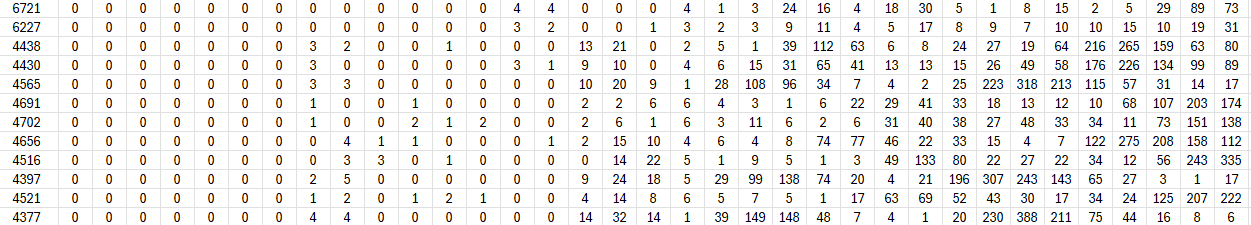}}
\caption{Temporal representation of message spread with 100-millisecond intervals}
\label{fig:time-spread}
\end{figure*}

\section{Proposed Modifications} \label{Proposed}
Removing duplicates without compromising resilience and message dissemination latency is crucial for improving GossipSub's effectiveness in handling large messages. Therefore, we propose targeted improvements to the current protocol operation. It is essential to highlight that the proposed improvements take effect only when the message size exceeds a specified threshold. For better understanding, we first highlight the protocol components responsible for performance deterioration and then suggest remedies to better cope with the shortcomings. Notably, the proposed modifications preserve the protocol's robustness against adversarial behavior.

\subsection{PREAMBLE for Quick Identification of Messages} \label{preamble}
Ignoring any delays related to message processing and similar factors, the message transfer time to a single peer can be simplified to $\tau_1^S \approx \frac{S}{R} + \tau_p$, with $S$, $R$, and $\tau_p$ being the message size, data rate, and average link latency, respectively. This implies that message transmission time over a 100 Mbps link with 100 ms latency jumps from $\tau_1^{10KB} \approx 100.8\ ms$ for a 10KB message to $\tau_1^{1MB} \approx 180\ ms$ for a 1MB message. This rise in message transmission time can be attributed to the larger message size. The transfer time to full-message mesh with $D = 8$ rises to $\tau_D^{1MB} \approx 740 ms$. Experiments reveal that the $\tau_D^{1MB}$ further rises noticeably, due to the congestion avoidance mechanisms available at the transport layer. For instance, consider spreading twelve messages (1MB each) in a network comprising 1500 peers, each having 100 Mbps bandwidth and each link having 100 ms latency. Fig. \ref{fig:time-spread} provides a temporal representation of this message spread. Each row corresponds to an individual message, with the first row showing the message spread of the earliest transferred message and the last row depicting the final message. The first column provides network-wide message dissemination time for each message, while the subsequent columns indicate the number of peers that received the message during each 100-millisecond interval. It takes approximately 900 to 2000 milliseconds for publishers to deliver messages to peers in their full-message mesh (with $D = 8$), with initial messages taking longer to complete due to a smaller $C_{wnd}$.

It is important to understand that peers are unaware of the messages they are receiving, and reception times are considerably higher for large messages. Therefore, it is essential to make receivers immediately aware of the key details of the messages they are receiving.
Rapid identification of incoming messages leads to mechanisms that can alleviate many redundant transmissions without compromising the robustness and latency of the network. To achieve this, we propose a short control message called PREAMBLE, which is transmitted immediately before every large message transmission. The PREAMBLE serves as a commitment from the sender, indicating that the promised data message will follow without delay. The included msgID and corresponding message length enable the receiver to quickly recognize and prepare for the upcoming message transfer. 
A receiving peer must immediately process a received PREAMBLE without waiting to download the entire message. This approach allows continuous maintenance of the 'ongoing\_receives' list at receiving peers. 
It is essential to highlight that an IDONTWANT message can also serve as a PREAMBLE, provided the peer sending IDONTWANT transfers the announced message without delay. However, this requires modifying the semantics of IDONTWANT to include a commitment for immediate message transfer. 
Similarly, prepending a PREAMBLE to the outgoing large message can eliminate the delay between PREAMBLE signaling and the corresponding message transfer. However, this approach introduces additional complexity,  as the multiplexer must be modified to immediately forward received PREAMBLEs to the upper layer while simultaneously continuing to receive remaining message payloads.
The Performance evaluations in this article use IDONTWANT messages as PREAMBLEs.

\subsection{IMRECEIVING Message to Minimize Duplicates} \label{IMRECEIVING}
For every message, a peer typically makes up to $D$ transmissions to contribute its fair share to the spread of messages. However, the fact that many recipients may have already received the message from other peers gives rise to the need for efficient message forwarding. Although the $D$-spread is associated with quicker dissemination and resilience against non-conforming peers, many potential solutions can still minimize redundant transmissions while preserving the GossipSub resilience. For instance, IDONTWANT announcements allow any node to notify its mesh members that it has already received the message, thereby preventing them from resending it. However, the use of IDONTWANT messages has an inherent limitation, i.e., a peer can send IDONTWANT announcements for a message only after it has received the entire message. Since transmitting a large message takes considerable time, other mesh members may also start transmitting the same message during this period. Therefore, the probability of simultaneously receiving the same message from multiple senders increases with the message size, compromising the effectiveness of IDONTWANT messages.

Upon receiving a PREAMBLE, a peer immediately learns the msgID and length of an incoming message, as detailed in section \ref{preamble}. In response, the peer must immediately request its remaining mesh members to refrain from resending the same message. 
To facilitate this, we propose a new control message called IMRECEIVING. IMRECEIVING announces the msgID and length (extracted from PREAMBLE) to mesh members, asking them to refrain from resending the announced message. 
IMRECEIVING addresses the limitation of IDONTWANT messages. For instance, consider a peer $X$ begins receiving a message $M$ at time $t_1$. It can transmit IDONTWANT only after receiving $M$, i.e., at time $t_1+\tau_D$. Therefore, it can not cancel any duplicate receptions of $M$ that start before $t1 + \tau_D + \tau_p$. In contrast, IMRECEIVING announcements for $M$ start at $t_1 + \Delta$, where $\Delta$ denotes PREAMBLE processing time and satisfies $\Delta \ll \tau_D$. As a result, peer $X$ can eliminate all duplicate receptions of $M$ that start after $t_1 + \Delta +\tau_p$, which noticeably reduces duplicates.

While IMRECEIVING messages can substantially minimize latency and bandwidth utilization, knowing the potential risks is crucial. A malicious user can exploit this approach by sending a PREAMBLE and never completing (or deliberately delaying) the promised message transfer, which can hinder the message propagation in the network. 
To mitigate this problem, we propose using a simple defense mechanism. 
Upon receiving a PREAMBLE from peer $X$, peer $Y$ uses the length field to conservatively estimate message transfer duration, ensuring sufficient time allocation for reception. If $X$ fails to complete the transfer during this duration, $Y$ penalizes $X$ through negative peer scoring.
In this case, message dissemination to $Y$ can follow a push-based or a pull-based strategy based on the application requirements. 
In a pull-based model, $Y$ issues an IWANT request to retrieve the message from mesh members ($Y$ can infer message availability at mesh members through IDONTWANT announcements).
In a push-based model, mesh members proactively send this message to $Y$ if they do not receive a corresponding IDONTWANT announcement from $Y$ within the estimated message transfer duration. 
To support this behavior, $Y$ may include the message transfer duration in the IMRECEIVING announcement. Alternatively, mesh members can independently derive a conservative estimate for duration based on the length field. 
In this article, we stick to a push-based operation, where mesh members proactively send the message to $Y$ if the estimated message transfer duration finishes without IDONTWANT notification from $Y$.
It is also important to note that using msgID and length together as a message descriptor in PREAMBLE and IMRECEIVING messages makes it difficult for a malicious peer to misrepresent the message length.

\begin{table*}[t]
  \caption{Simulation Scenarios}
  \centering
  \begin{tabular}{|c|c|c|c|c|c|}
    \hline
    \textbf{Experiments}& \textbf{No. of Nodes}& \textbf{No. of Publishers}& \textbf{Message Size (KB)}& \textbf{Inter-Message delay}& \textbf{Results}\\
    \hline
    Scenario 1 & 1500 & 12 & 200, 400, 600, ..., 1400 & 4 seconds & Fig. \ref{fig:idontwant_avg_dups}-(a), Fig. \ref{fig:inc_message_size}\\
    \hline
    Scenario 2 & 1500 & 22, 42, 62, ..., 142 & 50 & 50 milliseconds & Fig. \ref{fig:idontwant_avg_dups}-(b), Fig. \ref{fig:inc_nw_and_publishers}-(a,c,e)\\
    \hline
    Scenario 3 & 2000, 4000, ..., 12000 & 22 & 100 & 3 seconds & Fig. \ref{fig:idontwant_avg_dups}-(c), Fig. \ref{fig:inc_nw_and_publishers}-(b,d,f)\\
    \hline
  \end{tabular}
  \label{tab:Scenarios}
\end{table*}

\begin{table}[htb]
  \caption{Simulation Parameters}
  \centering
  \begin{tabular}{|c|c|c|c|}
    \hline
    \textbf{Parameter}& \textbf{Value}&\textbf{Parameter}&\textbf{Value}\\
    \hline
    $D$ & 8 & $D_{low}$ & 6\\
    \hline
    $D_{lazy}$ & 6 & $D_{high}$ & 12\\
    \hline
    $D_{out}$ & $\frac{D_{low}}{2}$ & gossipFactor & 0.05 \\
    \hline
    Heartbeat Interval & 700 ms & FloodPublish & false\\    
    \hline
    Muxer & yamux & Reduced Send ($K$) & $D_{low}-1$\\
    \hline
  \end{tabular}
  \label{tab:parameters}
\end{table}

\subsection{IWANT Message Improvements} \label{IHAVEIWANT}
In a GossipSub network, full-message meshes are usually sufficient for delivering messages to all subscribed peers, provided that they form a connected subgraph. Typically, a moderate mesh degree is adequate for maintaining a connected subgraph. However, in the presence of adversaries or network partitions, IHAVE messages ensure that any peer can request a missing message using an IWANT request. Another crucial role of IHAVE messages is to help distant peers expedite the retrieval of overdue messages through IWANT requests, indirectly lowering network-wide message dissemination latency. This arrangement works well for small messages. However, higher transmission times for large messages lead to two fundamental problems:
1) Responding to IWANT requests is optional in GossipSub v1.1. Therefore, peers typically make multiple IWANT requests to different hosts for the same message as a measure to ensure successful message reception. 
2) A peer remains unaware of ongoing message receptions until they are complete. During this interval, it may initiate multiple IWANT requests for these messages. 
Notably, most IWANT requests are serviced by early message receivers or by peers situated along the optimal forwarding paths. These unnecessary IWANT requests lead to increased message dissemination time and bandwidth utilization. 

PREAMBLE provides a simple remedy to this problem by notifying peers about messages they are receiving. As a result, peers refrain from sending IWANT requests for these messages. At the same time, restricting the number of outstanding IWANT requests for a large message to one can help minimize redundant transmissions. However, this adjustment requires that all peers immediately respond to incoming IWANT requests. It is important to note that some non-conforming peers may choose not to respond to incoming IWANT requests. Such peers can be penalized through negative peer scoring, and a new IWANT request can be made after the estimated message download duration finishes.

\subsection{Reduced Forwarding for a Manageable Workload} \label{ReducedSending}
It is common for slow peers to accumulate outgoing message queues, especially when handling large message transfers. This situation can lead to significant delays for subsequent outgoing messages. One possible solution to mitigate this problem is to lower the mesh degree $D$, which can help decrease the workload of slower peers by compromising the resilience of the GossipSub protocol. 
A better alternative is to preserve the recommended value for $D$ with a minor modification, i.e., on receiving a large message, a peer relays it to randomly selected $K \leq D$ mesh members and sends an IHAVE announcement to the remaining mesh members without waiting for the heartbeat interval. 
The issuance of IHAVEs announces immediate data availability to mesh members, thereby preserving the same level of redundancy in the presence of adversaries.
Consequently, remaining mesh members can fetch missing messages using IWANT requests without waiting for the heartbeat interval. Extensive simulations reveal that setting $K = D_{low} - 1$ yields good performance.

\section{Results and Discussions} \label{Results}
This section provides a comparative performance evaluation of the proposed schemes against GossipSup v1.2. To achieve this, we extend the nim-libp2p implementation \cite{status2023nim-libp2p} by incorporating support for PREAMBLE. This support enables the use of IMRECEIVING message functionality and IWANT improvements. A hybrid protocol (GossipSub v1.4) \cite{gossipsub1_4poc2025} that combines the functionalities and benefits of IMRECEIVING and IWANT improvements is also used. It is important to note that the GossipSub v1.4 specification proposal is already under review for standardization \cite{gossipsub1_42024}. Furthermore, we also report results from the reduced forwarding scheme, where a peer forwards messages to only $D_{low}-1$ peers and sends IHAVE announcements to the remaining mesh members.
We evaluate performance using three key metrics: network-wide message dissemination latency, bandwidth utilization, and average duplicates. Network-wide message dissemination latency ($L_{cov}^{100}$) measures the time it takes for a message to reach the entire network. For each experiment, we transmit multiple messages in the network. The first two messages are warmup messages (aimed to raise $C_{wnd}$), and we exclude them from all computations except bandwidth utilization. $L_{cov}^{100}$ averages the network-wide message dissemination time for the remaining messages. Bandwidth utilization ($B_N$) provides the total traffic volume, including control traffic and actual data transmissions. A peer may receive multiple copies of any transmitted message from mesh members. Excluding the first received message, all copies are duplicates. We compute average duplicates received by a peer as $\bar{d} = \frac{1}{N M} \sum_{j=1}^{M} \sum_{i=1}^{N} d_{i,j}$ where $N$ and $M$ denote the number of peers and the number of transmitted messages, respectively, and $d_{i,j}$ represents the number of duplicates received by peer $i$ for message $j$.

\begin{figure*}[htb]
    \centering
    \subfigure[Increasing message size (200KB-1.4MB message size)]{
        \includegraphics[width=0.45\textwidth]{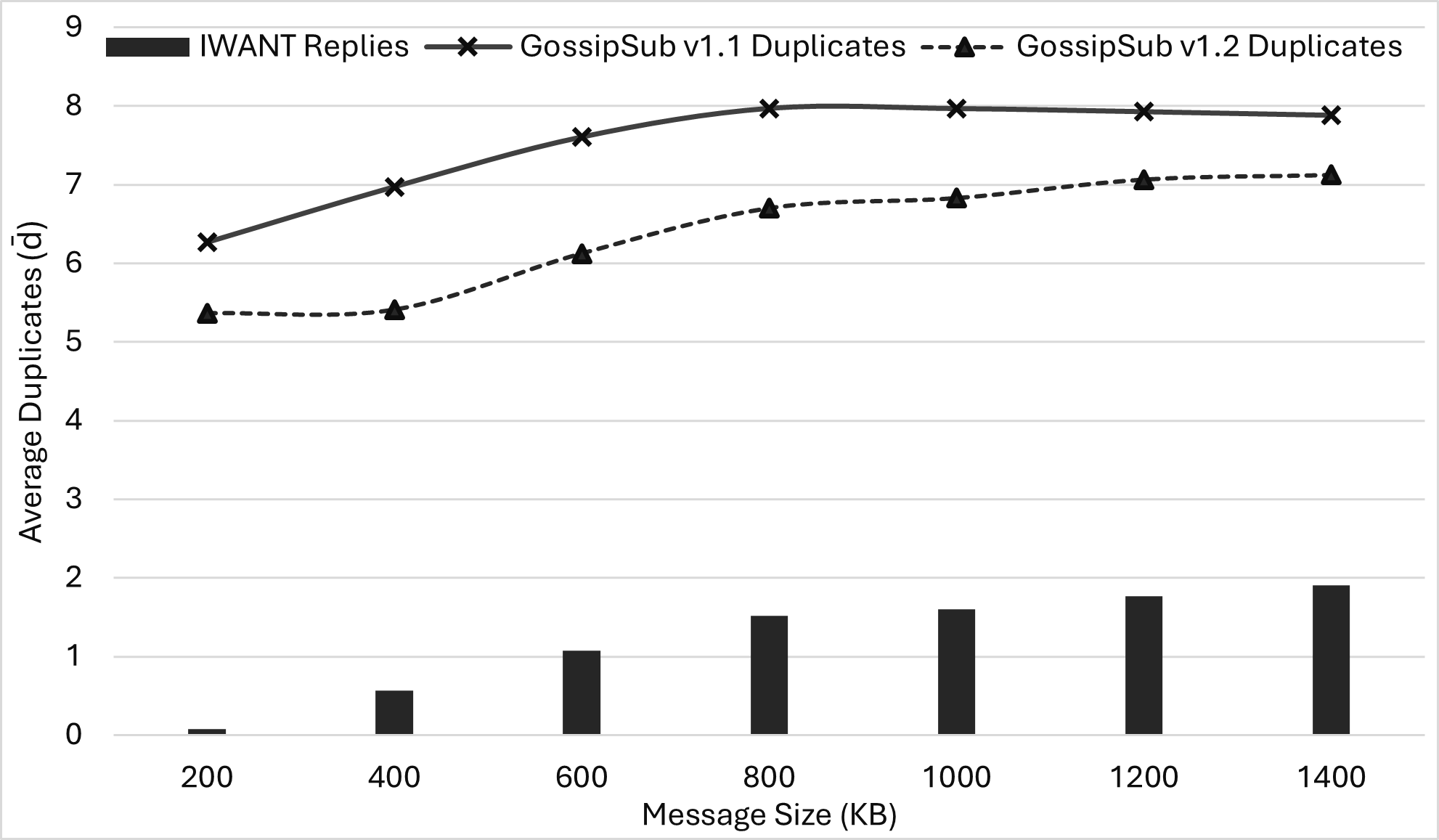}
    }
    \subfigure[Increasing number of publishers (50KB message size)]{
        \includegraphics[width=0.45\textwidth]{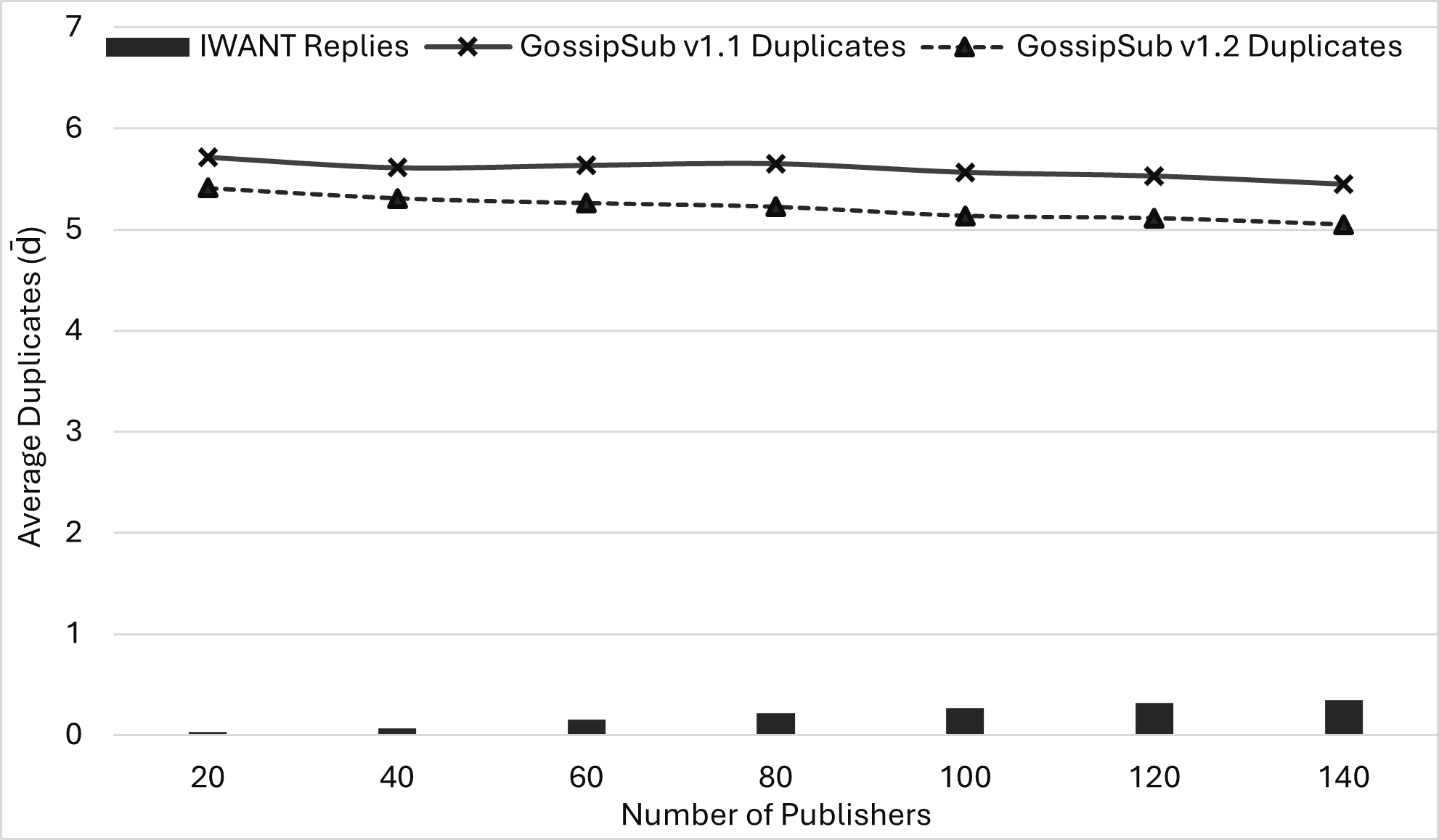}
    }
    \subfigure[Increasing number of nodes (100KB message size)]{
        \includegraphics[width=0.45\linewidth]{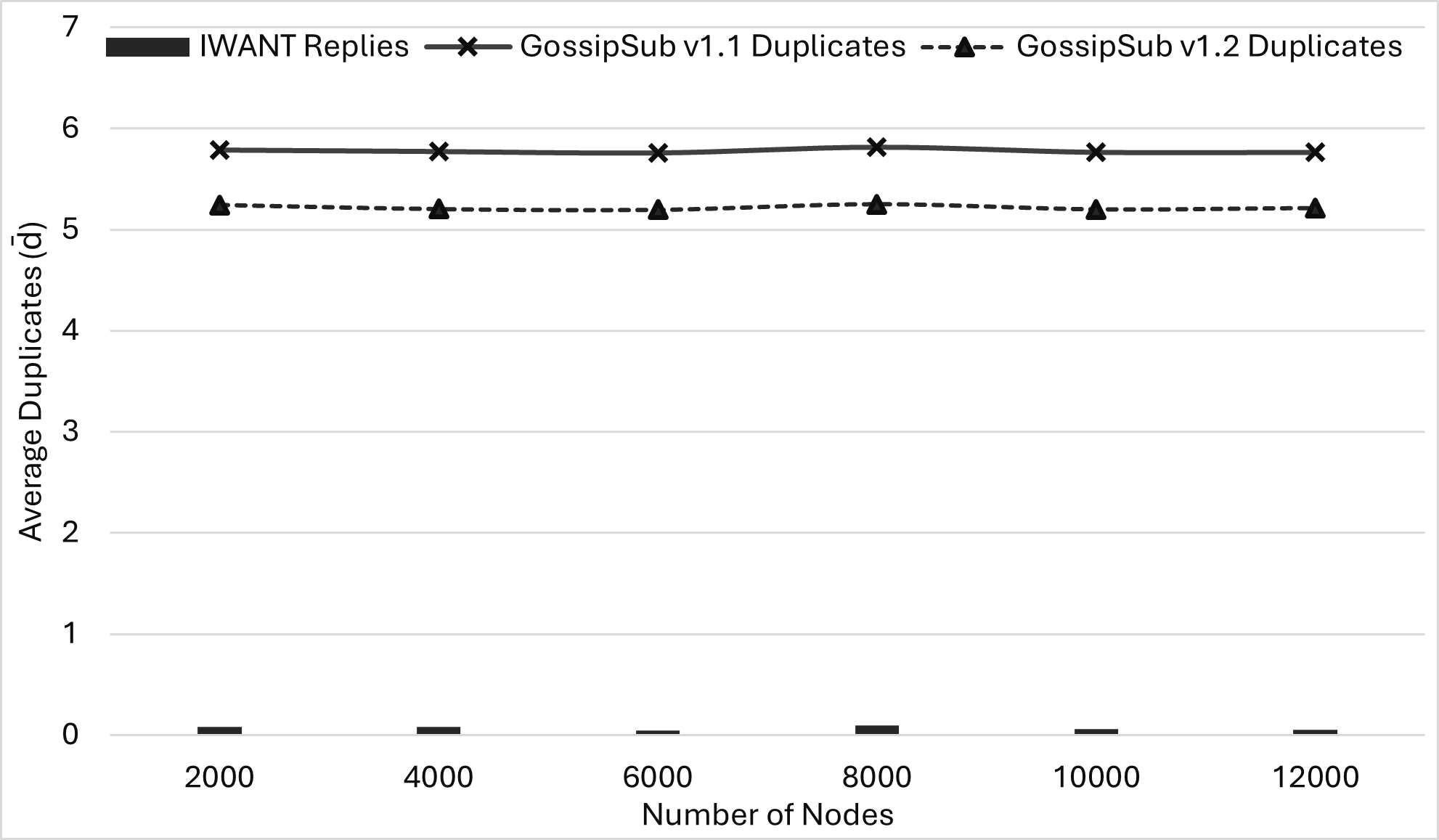}
    }    
    \caption{Average number of duplicates ($\bar{d}$): impact of IDONTWANT and IWANT-reply messages}
    \label{fig:idontwant_avg_dups}
\end{figure*}

Three simulation scenarios are considered: 1) The number of publishers and the network size remain constant while the message size gradually increases. 2) The number of nodes and message size remain constant while the number of publishers gradually increases. 3) The number of publishers and message size are kept constant while the network size gradually increases.
In all experiments, we transmit multiple messages such that every publisher sends exactly one message to the network. After a publisher transmits a message, each subsequent publisher waits for a specified interval (inter-message delay) before sending the next message. Rotating publishers ensures that every message traverses a different path, which helps achieve fair performance evaluation. 
On the other hand, changing inter-message delays allows for creating varying traffic patterns. A shorter delay implies more messages can be in transit simultaneously, which helps evaluate performance against large message counts. A longer delay ensures every message is fully disseminated before introducing a new message. Similarly, increasing message size stresses the network. As a result, we evaluate performance across a broader range of use cases. 
A detailed description of the simulation scenarios is presented in Table \ref{tab:Scenarios}. The experiments are conducted using the shadow simulator \cite{shadow2023simulator}. We uniformly set peer bandwidths and link latencies between 50-150 Mbps and 40-130 milliseconds in five variations. The publishers are uniformly selected from all bandwidth classes. GossipSub-related simulation parameters are provided in Table \ref{tab:parameters}.

\begin{figure*}[htb]
    \centering
    \subfigure[Message dissemination latency ($L_{cov}^{100}$)]{
        \includegraphics[width=0.45\textwidth]{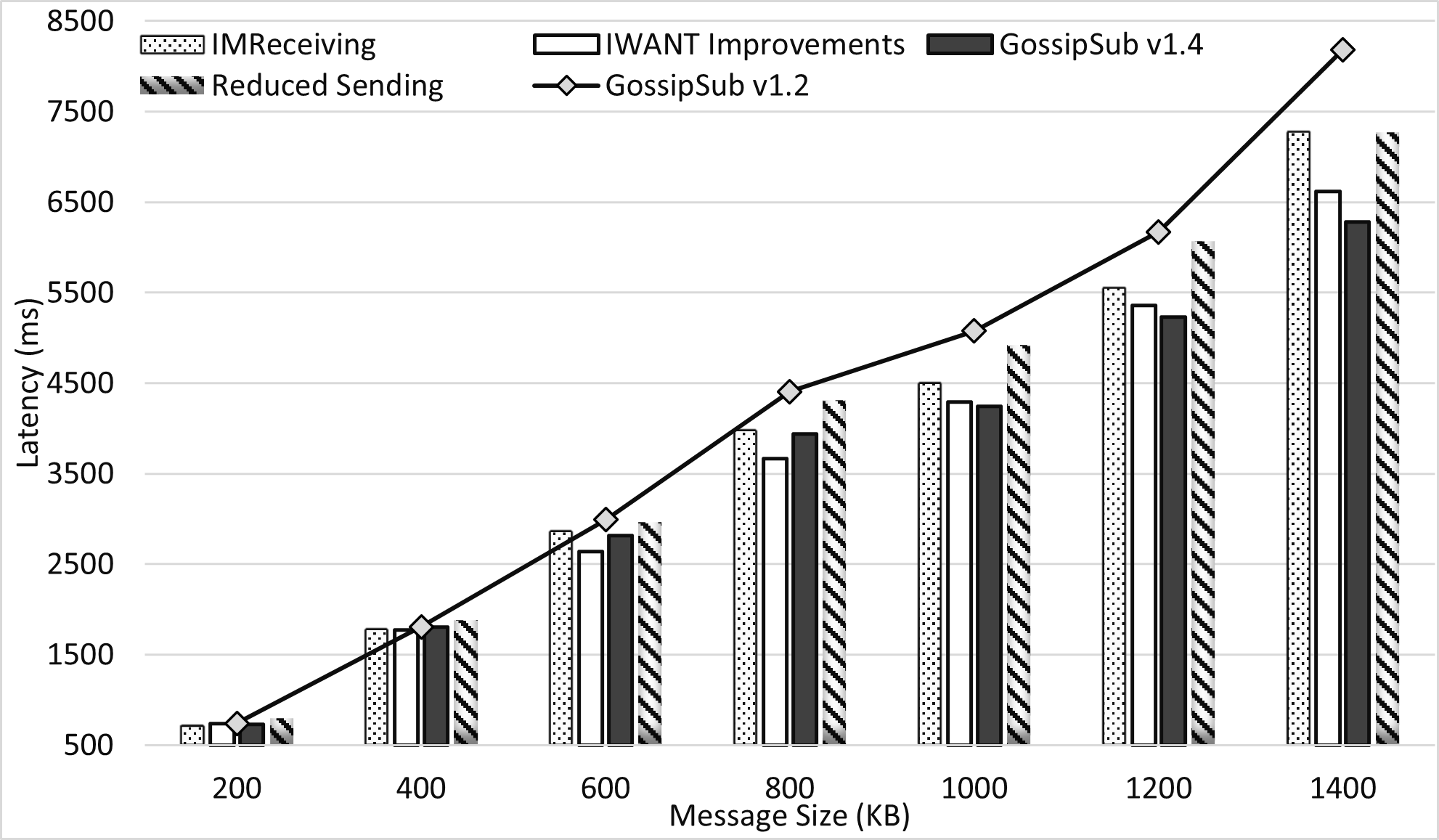}
    }
    \subfigure[Network-wide bandwidth utilization ($B_N$)]{
        \includegraphics[width=0.45\linewidth]{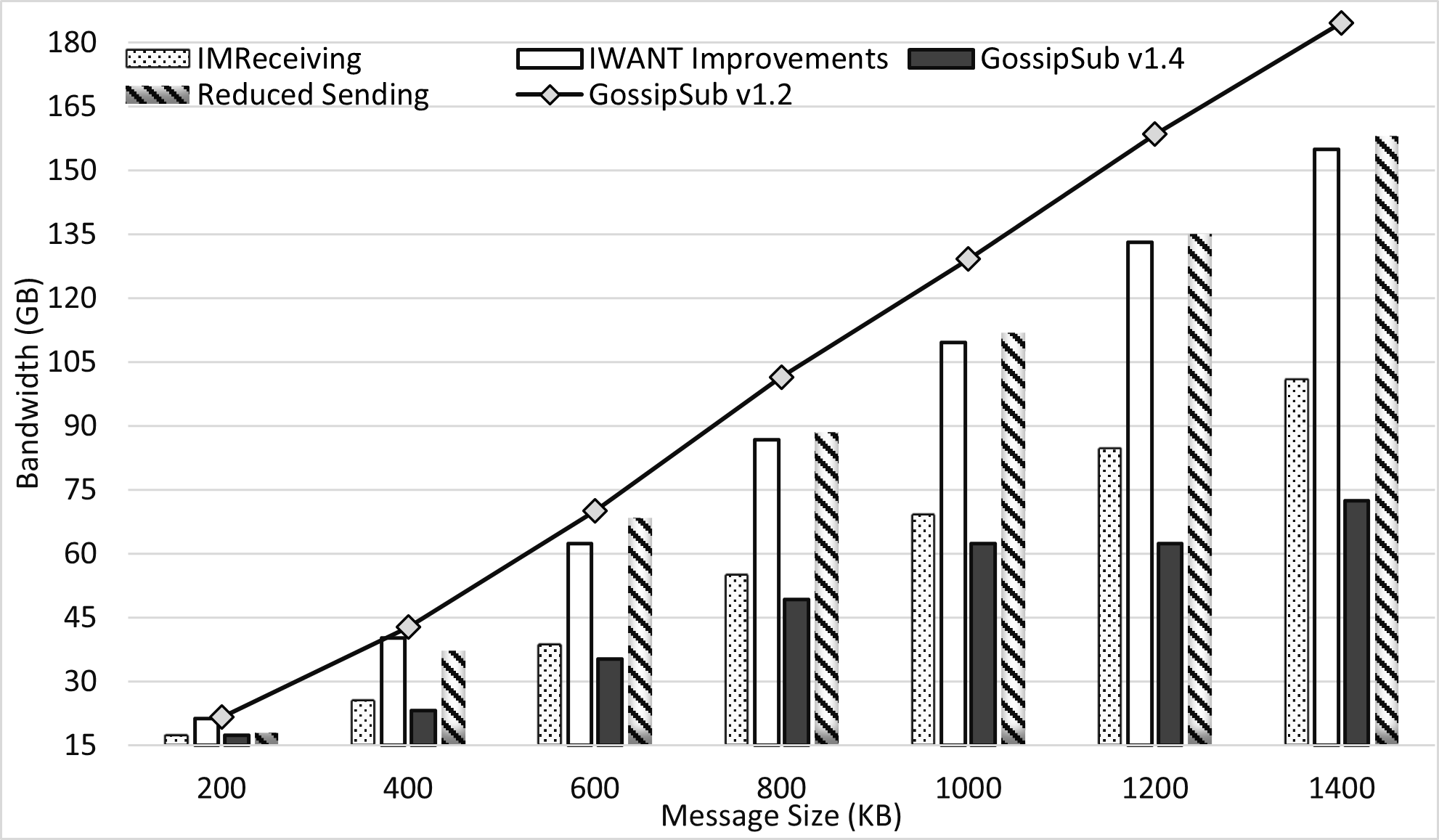}
    }
    \subfigure[Total number of IWANT requests]{
        \includegraphics[width=0.45\textwidth]{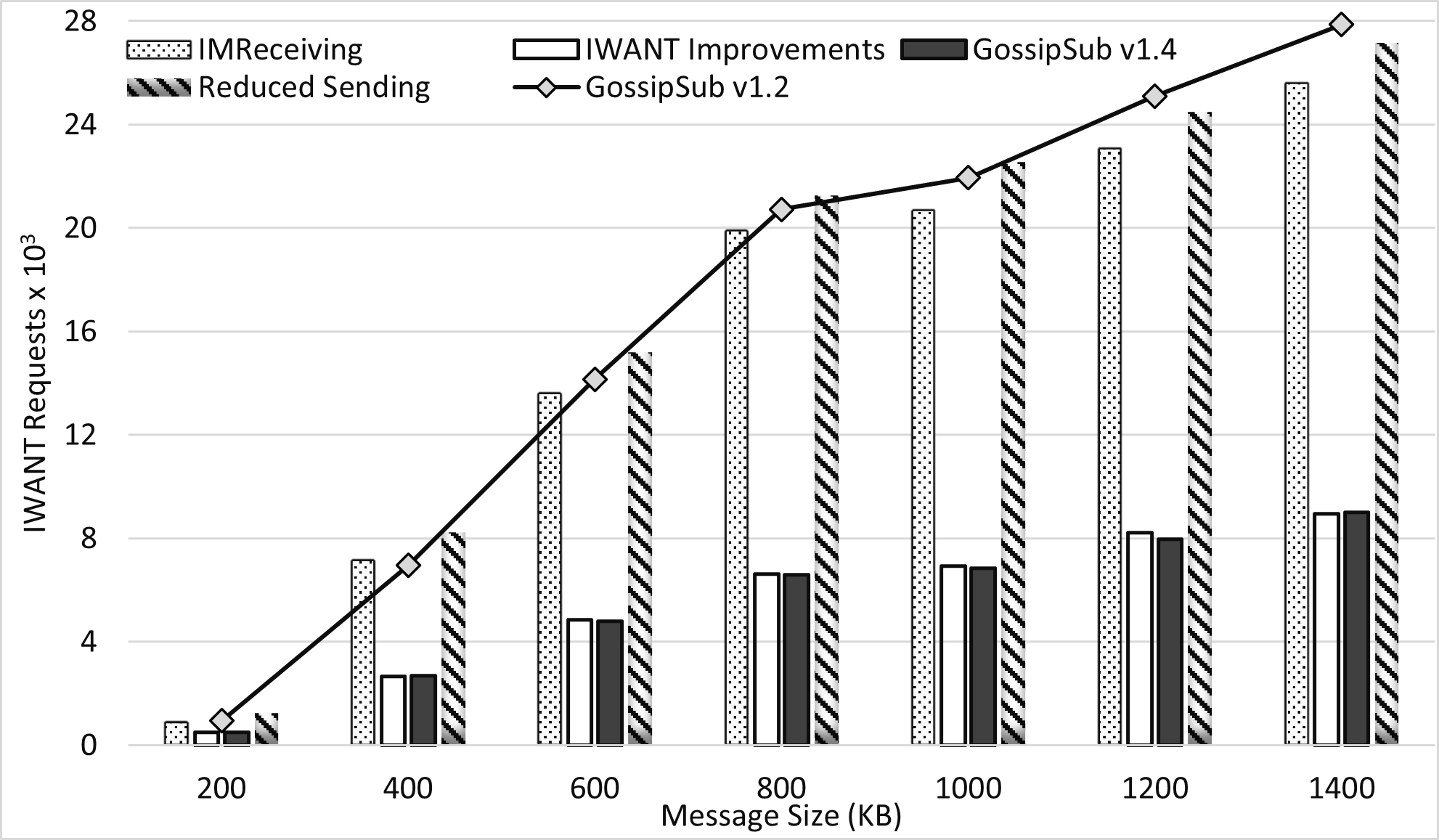}
    }
    \subfigure[Average number of duplicates ($\bar{d}$)]{
        \includegraphics[width=0.45\textwidth]{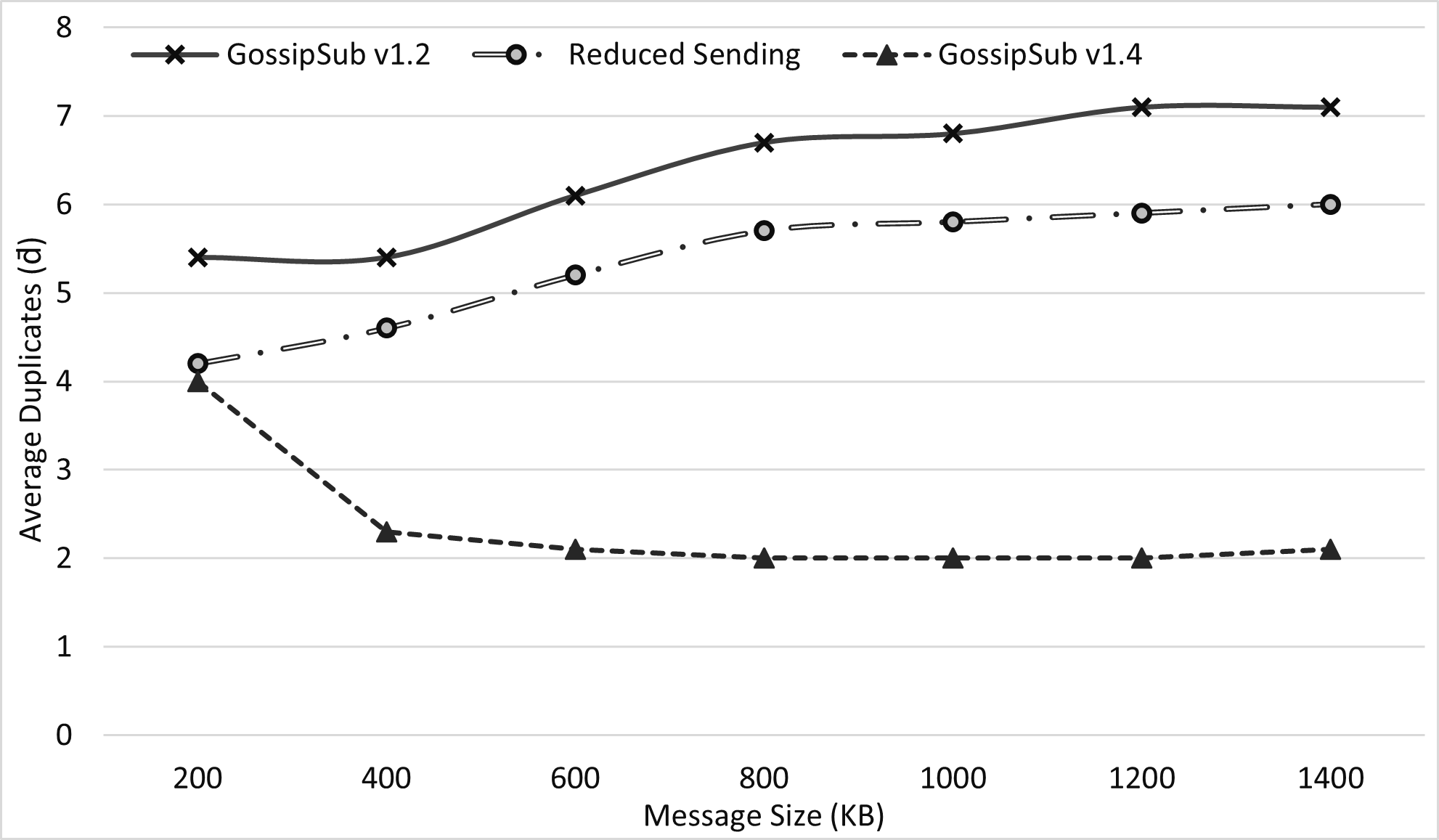}
    }
    \caption{Performance against increasing message size: latency ($L_{cov}^{100}$), bandwidth ($B_N$), average duplicates ($\bar{d}$)}
    \label{fig:inc_message_size}
\end{figure*}

\begin{figure*}[htb]
    \centering
    \subfigure[Message dissemination latency ($L_{cov}^{100}$)]{
        \includegraphics[width=0.46\textwidth]{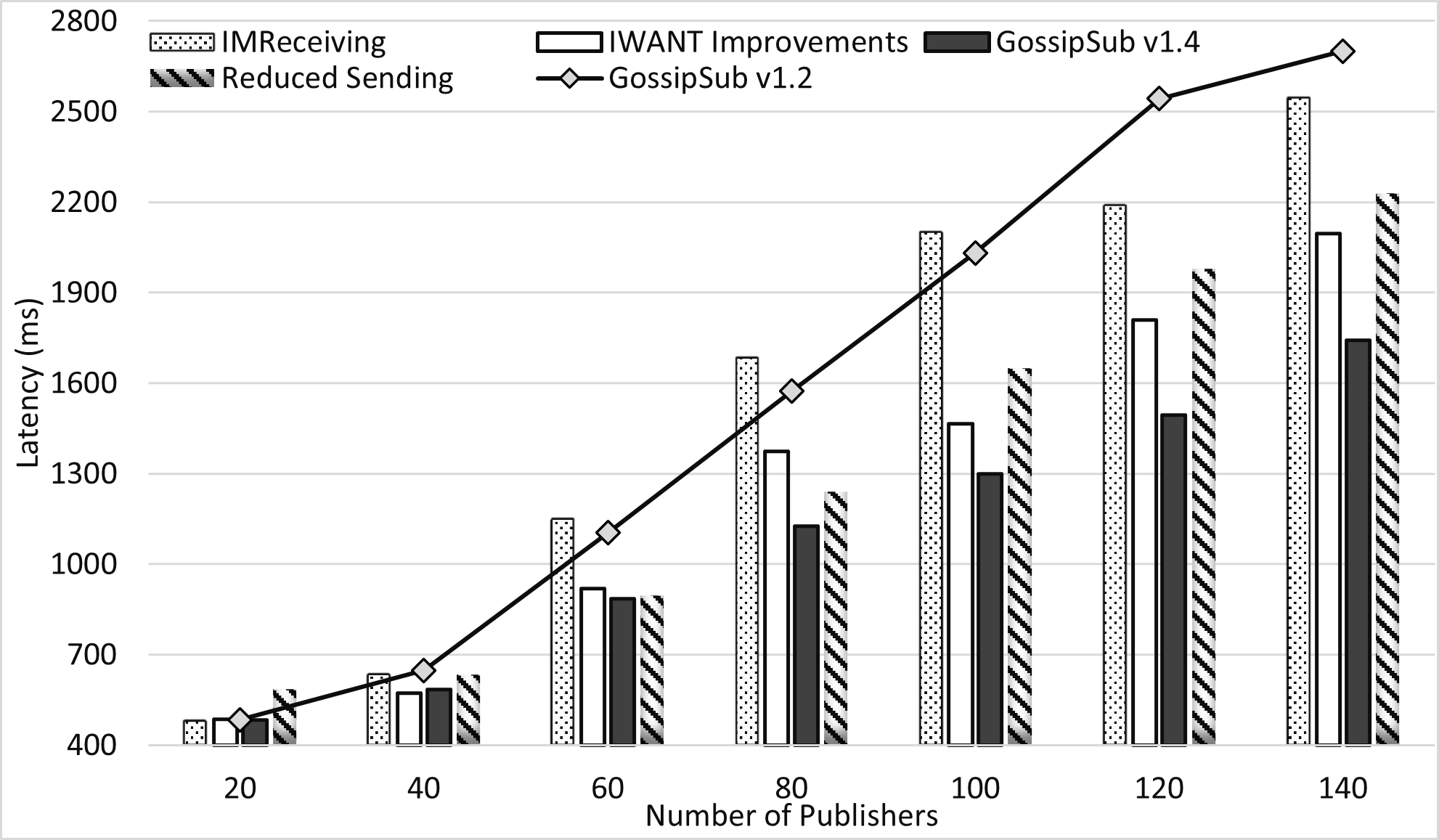}
    }
    \subfigure[Message dissemination latency ($L_{cov}^{100}$)]{
        \includegraphics[width=0.46\linewidth]{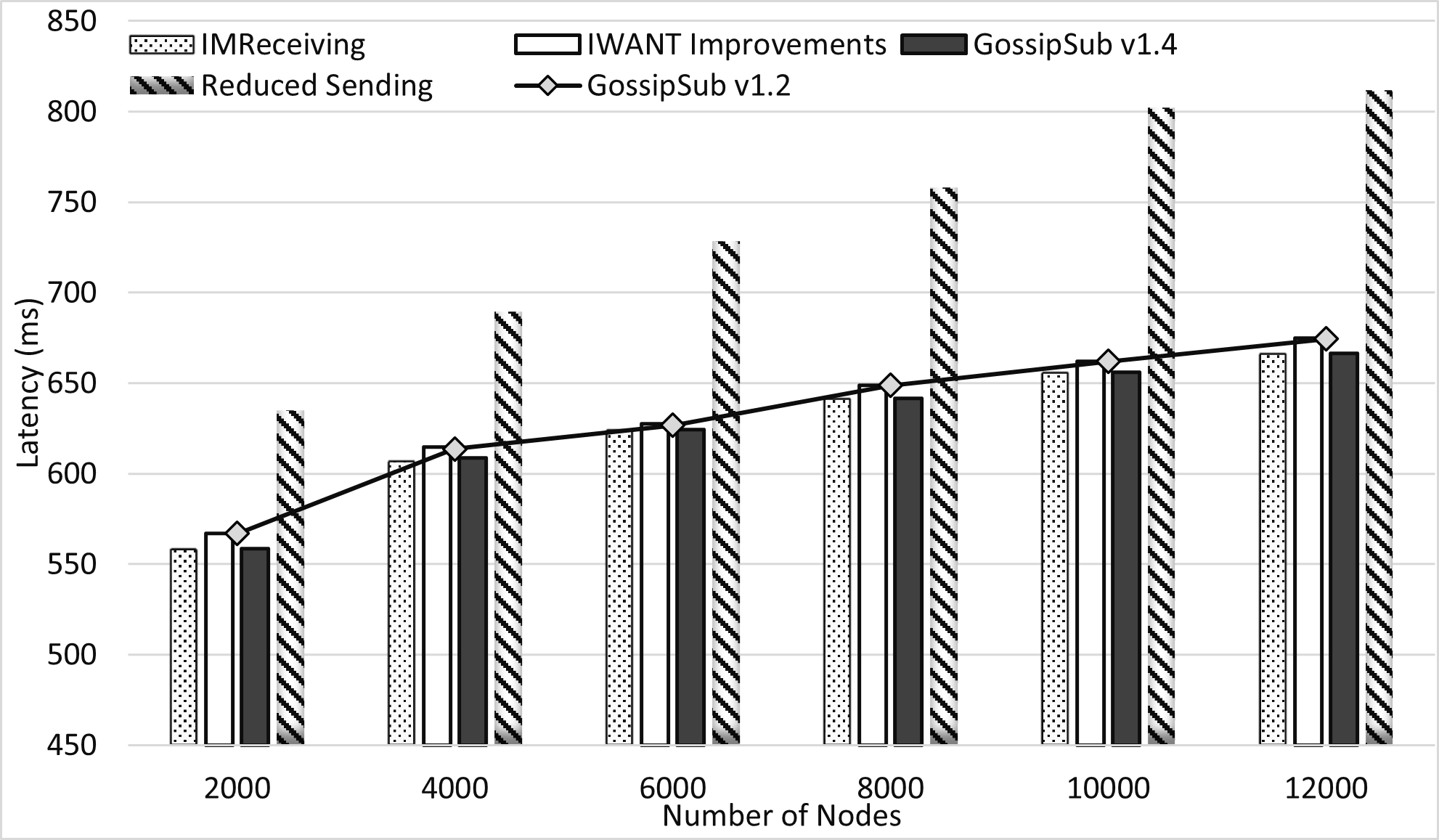}
    }
    \subfigure[Network-wide bandwidth utilization ($B_N$)]{
        \includegraphics[width=0.46\textwidth]{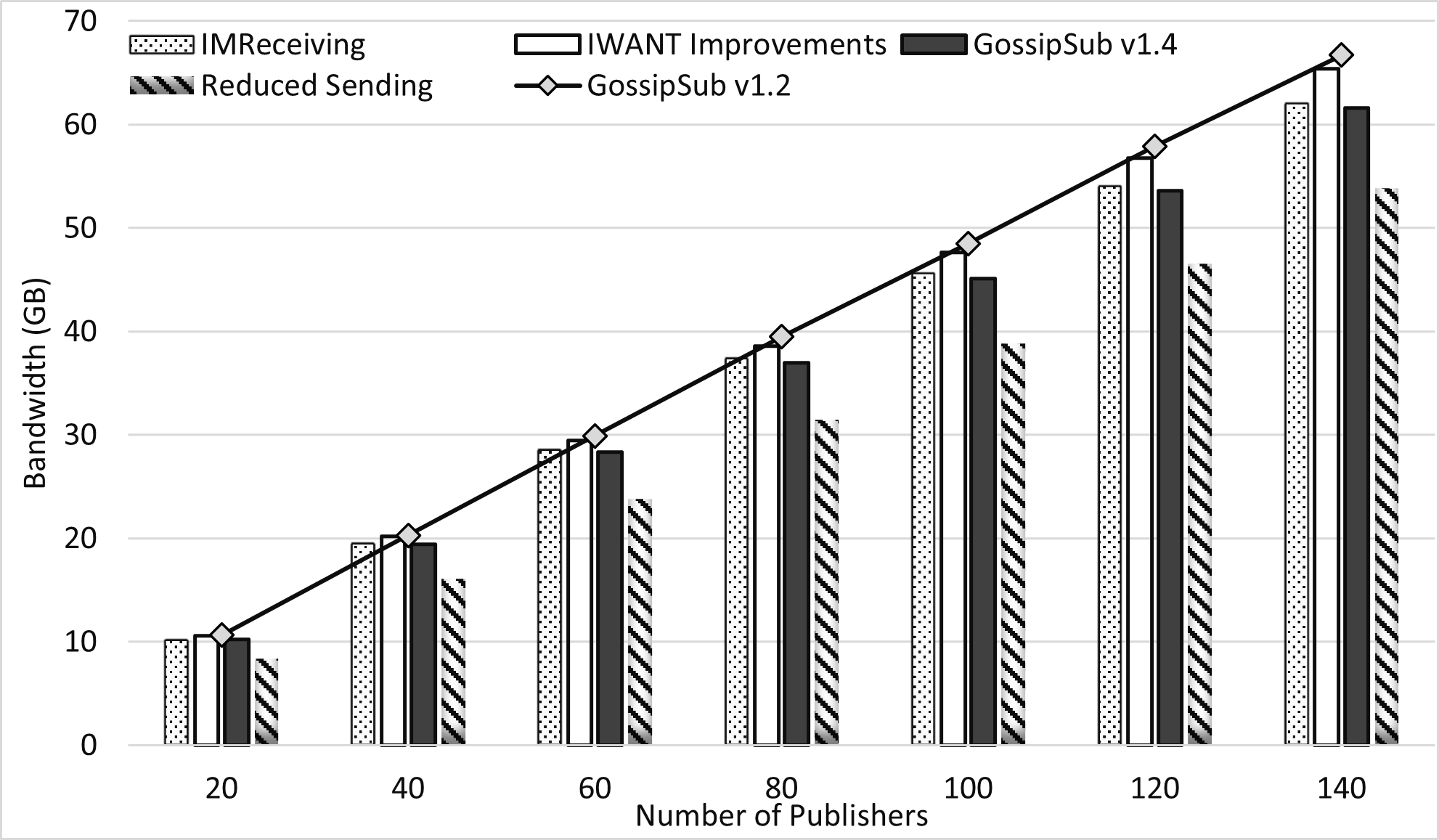}
    }
    \subfigure[Network-wide bandwidth utilization ($B_N$)]{
        \includegraphics[width=0.46\textwidth]{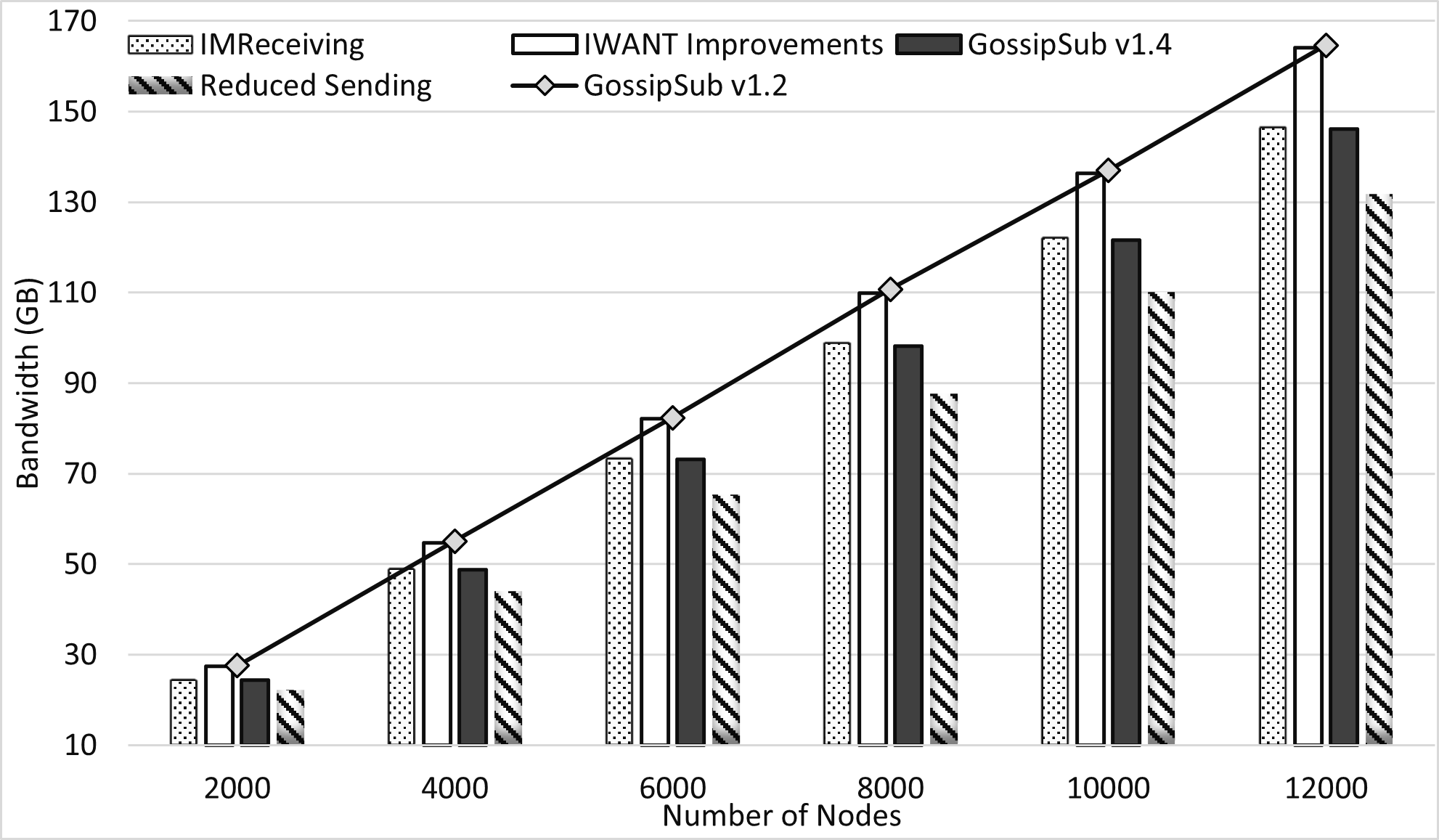}
    }
    \subfigure[Average number of duplicates ($\bar{d}$)]{
        \includegraphics[width=0.46\textwidth]{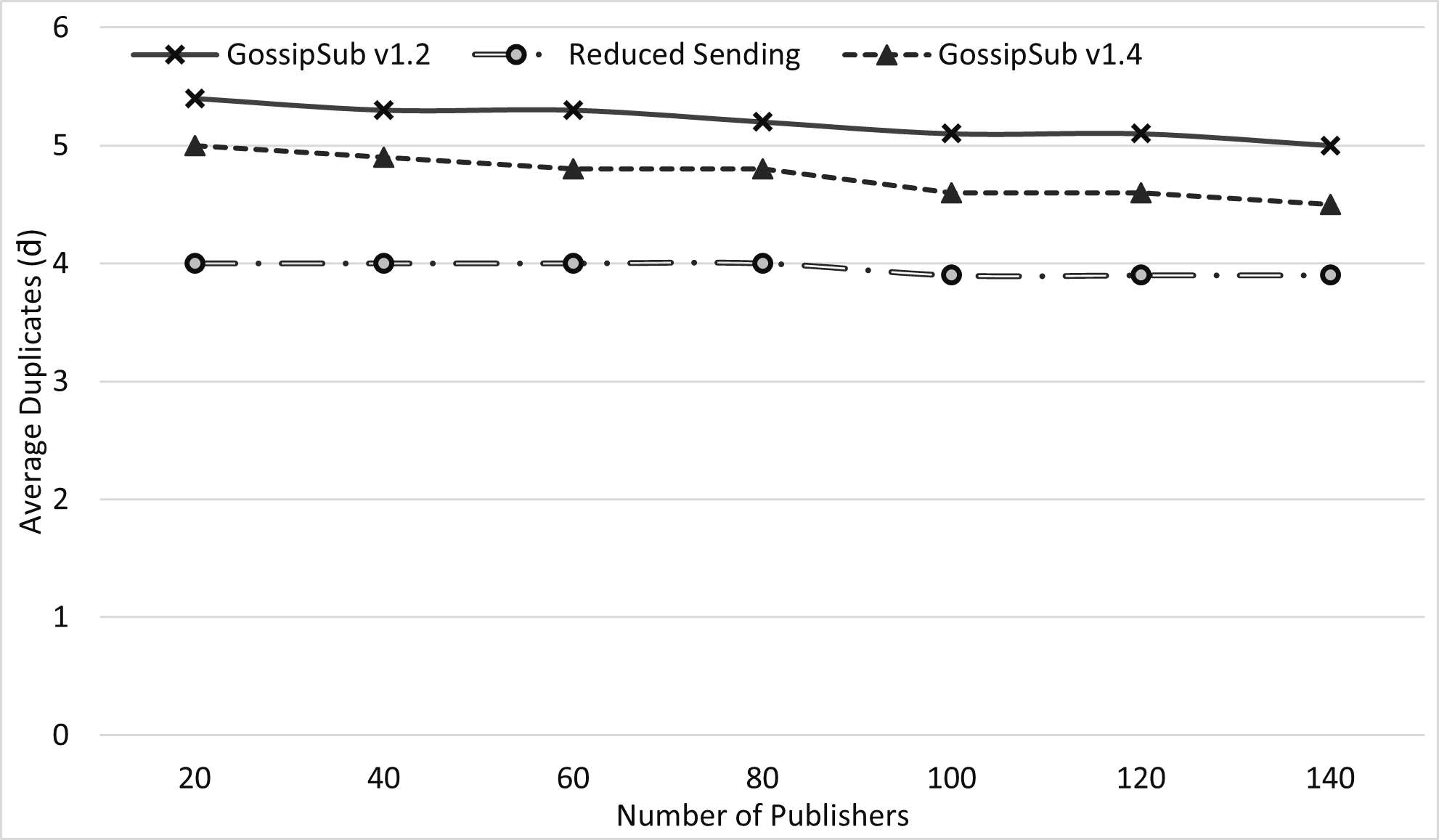}
    }
    \subfigure[Average number of duplicates ($\bar{d}$)]{
        \includegraphics[width=0.46\textwidth]{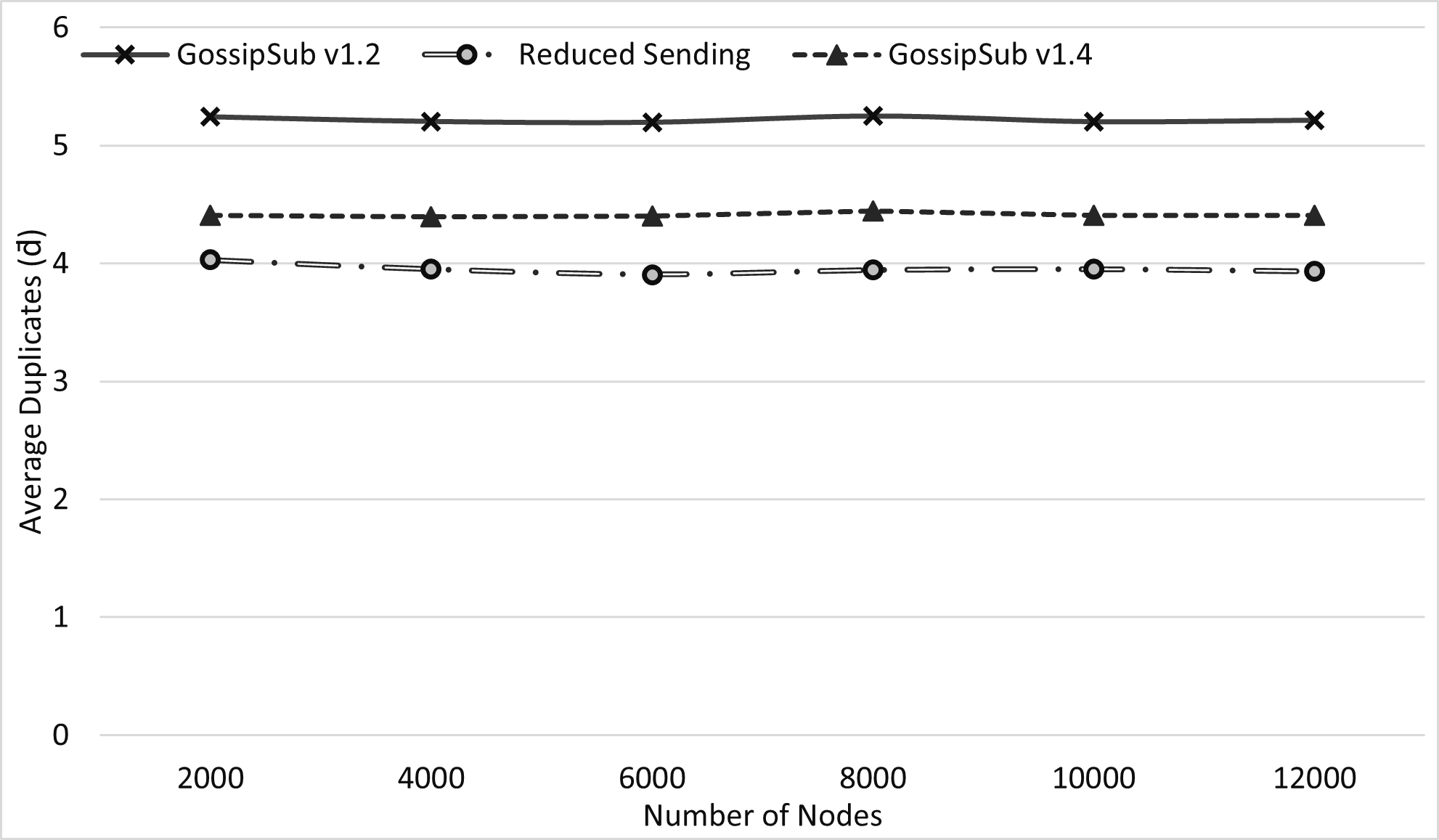}
    }
    
    \caption{Increasing number of publishers (a,c,e) and increasing number of nodes (b,d,f): latency ($L_{cov}^{100}$), bandwidth ($B_N$), average duplicates ($\bar{d}$)}
    \label{fig:inc_nw_and_publishers}
\end{figure*}

We begin by discussing average duplicates ($\bar{d}$). A network comprising $N$ peers, each with a degree $D$, has a total of $\frac {N \times D}{2}$ edges (links), as every link connects two peers. Assuming that a message traverses every link exactly once, network-wide dissemination requires at least $\frac {N \times D}{2}$ transmissions. Only $N-1$ transmissions are necessary for delivering a message to all peers. As a result, we get $\frac {N \times D}{2} -(N-1)$ duplicates in the network. We can simplify average duplicates received by a single peer to $\bar{d}_{min} \approx \frac{D}{2}-1$. Here, $\bar{d}_{min}$ represents the lower bound on average duplicates because we assume that the send and receive operations are mutually exclusive. This assumption requires that message transmission times (and link latencies) are so small that no two peers simultaneously transmit the same message to each other.
However, a large message can noticeably increase the contention interval (message transmission time). An extended contention interval raises the likelihood that many peers will simultaneously transmit the same message to each other. Authors in \cite{savolainen2020streamr} explore the upper bound ($\bar{d}_{max}$) on average duplicates. They argue that a node can forward a received message to a maximum of $D-1$ peers while the original publisher sends it to $D$ peers. As a result, we get $N(D-1)+1$ transmissions in the network. Out of these, $N(D-1)+1-(N-1)$ messages are duplicates, which simplifies the upper bound on average duplicates received by each peer to $\bar{d}_{max} \approx D-2$. This rise in average duplicates indicates the impact of increasing message size on duplicates. 
It is important to note that the number of IWANT replies also contributes to duplicate transmissions. However, it is not trivial to account for IWANT replies in $\bar{d}$ computations above.

Fig. \ref{fig:idontwant_avg_dups} compares average duplicates in GossipSub v1.1 and v1.2, which also includes duplicates contributed by IWANT replies. Notably, the number of IWANT replies is similar in both protocol versions. The average duplicates received by a peer remain below $D-2$ for 50KB and 100KB messages, as depicted in Fig. \ref{fig:idontwant_avg_dups}(b)-(c). These duplicates also include a small number of IWANT replies. The share of IWANT replies in $\bar{d}$ is also illustrated in the figure. On average, the share of IWANT replies accounts for less than 2\% of the average duplicates. However, this share rises to nearly 7\% when the number of publishers increases substantially, as large message counts lead to bigger outgoing message queue sizes. Consequently, peers generate a relatively higher number of IWANT requests.
On the other hand, Fig. \ref{fig:idontwant_avg_dups}(a) reports nearly $D$ duplicates for messages exceeding 600KB in size. This rise in average duplicates can be attributed to two factors: 1) large message sizes lead to longer contention times. As a result, many peers simultaneously start transmitting the same message to each other. 2) Longer transmission time solicits more IWANT requests. As a result, IWANT replies contribute 24\% duplicates for messages exceeding 1MB (nearly two duplicates per peer for every message). IDONTWANT messages in GossipSub v1.2 reduce average duplicates by up to 22\%. However, this reduction is less than expected because a peer can send IDONTWANT announcements only after downloading the entire message. For a prolonged download time, the receiver will likely begin receiving the same message from multiple mesh members, which undermines the effectiveness of IDONTWANT messages. Our previous findings report higher bandwidth preservation by IDONTWANT messages in homogeneous networks \cite{farooq2025staggering} because more peers receive messages within similar time intervals, allowing many of their transmitted IDONTWANTs to be processed in time. However, this effectiveness lowers in more realistic network conditions.

We now evaluate and compare the effects of proposed performance improvements against GossipSub v1.2. We use IDONTWANT messages as PREAMBLE to take advantage of IMRECEIVING and IWANT improvements. IMRECEIVING messages can minimize duplicates caused by the $D$-spread, and IWANT improvements aim to reduce the number of IWANT requests in the network. 
The merged approach (GossipSub v1.4) combines IMRECEIVING functionality with IWANT improvements to maximize performance gains. Finally, the reduced message sending involves transmitting a message to only $D_{low}-1$ peers and forwarding IHAVE announcements to the remaining mesh members. Lowering $D$ can minimize duplicates, while any missing peer can still use IWANT requests to download unreceived messages at the cost of an additional round-trip time. 
Fig. \ref{fig:inc_message_size}-(c) illustrates that IWANT improvements successfully eliminate many IWANT requests for messages above 200KB. Similarly, IMRECEIVING messages also successfully eliminate many duplicate transmissions to achieve a noticeable bandwidth reduction in Fig. \ref{fig:inc_message_size}-(b). As a result, the proposed GossipSub v1.4 achieves up to 45\% reduction in bandwidth utilization for 400KB messages. The reduction reaches 61\% for messages bigger than 1MB in size. As a result, GossipSub v1.4 brings the average number of duplicate messages per peer to approximately two for messages bigger than 400KB in size. On the other hand, the reduced message-sending approach lowers the average number of duplicate messages to approximately $D_{low}$, achieving roughly a 15\% decrease in bandwidth utilization. Reducing bandwidth utilization also helps minimize network-wide message dissemination latency by lowering outgoing message queue sizes on the peers at optimal or near-optimal paths. Therefore, a noticeable reduction in latency is expected for high traffic volumes. Proposed GossipSub v1.4 reflects this by achieving up to 23\% reduction in latency for large messages. It is also important to note that a higher mesh degree is typically associated with lower network-wide message dissemination times because it increases the probability that messages traverse optimal or near-optimal paths. As a result, up to 7\% higher latency is seen for a reduced message-sending approach for messages under 400KB in size. However, this approach yields a modest latency decrease (up to 11\%) as message sizes increase. However, the proposed GossipSub v1.4 noticeably outperforms GossipSub v1.2 and other methods in terms of latency and bandwidth utilization. 

Talking about smaller message sizes, IWANT improvements do not yield much impact on bandwidth utilization because IWANT replies have a minor share in average duplicates, as depicted in Fig. \ref{fig:inc_nw_and_publishers}(c)-(d). However, lowering duplicates through the reduced sending and IMRECEIVING messages is still possible. It is important to note that reduced sending can lead to higher bandwidth savings, but it may also increase message dissemination latency. This happens because lowering $D$ also decreases the likelihood of disseminating messages through optimal or near-optimal paths. Fig. \ref{fig:inc_nw_and_publishers}(b),(d) demonstrate that the reduced sending approach lowers bandwidth utilization by roughly 20\%, at the cost of a 20\% hike in message dissemination latency. In contrast, GossipSub v1.4 achieves approximately 11\% reduction in bandwidth utilization without compromising message dissemination latency. It is important to note that reducing duplicates in the presence of high message counts can help lower message dissemination latency by reducing workload at optimal/near-optimal path peers. This is why proposed GossipSub v1.4 yields up to 35\% reduction in message dissemination latency, whereas the reduced sending approach also lowers latency by roughly 18\%.

\section{Conclusion} \label{Conclusion}
This article investigates the pressing issue of high bandwidth utilization and a considerable rise in network-wide dissemination times when handling large messages. A key finding is that GossipSub does not consider the impact of message sizes when making forwarding decisions. Large messages lead to remarkably high mesh transmission times that increase the probability of peers simultaneously sending the same messages to each other. Higher mesh transmission times also cause a higher number of IWANT requests, resulting in considerable store-and-forward delays accumulating at each hop. As a result, duplicate messages consume a significant share of the available bandwidth. These duplicates have minimal impact on message dissemination latency for low traffic volumes. However, when traffic volume increases, outgoing message queue sizes rise, adding additional queuing delays to messages. It is important to note that minimizing workload at optimal path peers can significantly reduce message dissemination time by lowering the queuing and mesh transmission delays. That is why the IWANT message improvements lead to noticeable reductions in message dissemination latency. On the other hand, IMRECEIVING messages significantly reduce the number of duplicates, but this reduction yields a relatively lower impact on overall message dissemination latency. However, IMRECEIVING messages are generated at the start of message download. Compared to IDONTWANT messages, these quicker mesh notifications significantly reduce duplicates. It is important to note that messages like IMRECEIVING, IDONTWANT, and PREAMBLE must be transferred immediately. However, the lack of prioritization mechanisms at the transport layer can delay the transmission of these messages. The performance evaluations in this work are conducted without any transport layer prioritization. Having a priority system, such as QUIC stream prioritization, can greatly improve the effectiveness of these messages. 

The performance evaluations in this work assume a trusted environment and more generic use cases. Future work will investigate the robustness of the proposed GossipSub v1.4 protocol under adversarial conditions and evaluate its performance on Ethereum testnets.

\bibliographystyle{IEEEtran}
\bibliography{mybib}

\end{document}